\documentclass[twocolumn,aps,prx,longbibliography,amsmath,amssymb,floatfix,superscriptaddress,showkeys]{revtex4-2}

\usepackage{graphicx,xcolor}
\definecolor{lightgray}{rgb}{0.8,0.8,0.8}
\usepackage{booktabs,rotating} 

\renewcommand{\ol}[1]{\overline{#1}}
\newcommand{\rvec}{\mathbf{r}}
\newcommand{\tvec}{\mathbf{t}}
\newcommand{\n}{\mathbf{n}}
\newcommand{\bvec}{\mathbf{b}}
\newcommand{\e}{\mathbf{e}} 
\newcommand{\f}{\mathbf{f}} 
\newcommand{\eff}{\mathrm{eff}}

\newcommand{\micron}{\ensuremath{\mu\mathrm{m}}}

\newcommand{\pN}{\mathrm{pN}}
\newcommand{\s}{\mathrm{s}}

\newcommand{\Hz}{\mathrm{Hz}}
\newcommand{\nm}{\mathrm{nm}}
\newcommand{\fN}{\mathrm{fN}}
\newcommand{\Fhead}{F_\text{head}}
\newcommand{\rhoDMT}{\rho_\mathrm{DMT}}
\DeclareMathOperator{\sign}{sign}
\newcommand{\hide}[1]{}

\usepackage{xurl}
\usepackage[
  colorlinks=false,
  linkbordercolor=blue,
  citebordercolor=red,
  urlbordercolor=cyan,
  bookmarksopen=true,
  pdfauthor={Maximilian Kotz},
  pdftitle={Axonemal bending stiffness of Chlamydomonas cilia implies highly collective dynein motor dynamics}
]{hyperref}

\begin{document}
\title{Axonemal bending stiffness of \textit{Chlamydomonas} cilia implies\\ single-motor forces above $5\,\pN$}

\author{Maximilian Kotz}
\affiliation{Cluster of Excellence Physics of Life, TU Dresden, Dresden, Germany}
\author{Veikko F. Geyer}
\affiliation{B CUBE, TU Dresden, Dresden, Germany}
\author{Benjamin M. Friedrich}
\affiliation{Cluster of Excellence Physics of Life, TU Dresden, Dresden, Germany}

\email{benjamin.m.friedrich@tu-dresden.de}
\date{September 30, 2026}

\begin{abstract} 
The regular bending waves of cilia and flagella provide an iconic model system for the collective dynamics of molecular motors.
The known regular arrangement of dynein motors in a cilium's axoneme allows to connect mesoscopic cilia properties, such as axonemal bending stiffness, 
to microscopic motor properties, such as the force exerted by an individual motor.
We estimate the active force generated by the collection of dynein molecular motors in \textit{Chlamydomonas} axonemes using previous estimates of its bending stiffness. 
Divided by the maximal possible number of active motor heads, 
this provides a lower bound of ${>}10\,\pN$ for the peak force generated by a motor head, 
which exceeds typical stall forces ${<}5\,\pN$ of molecular motors.
This discrepancy suggests that either the bending stiffness of microtubules and axonemes was previously overestimated, 
or that collective force generation in dense motor arrays can surpass the sum of expected contributions of individual motors. 
\end{abstract}


\maketitle

\begin{figure*}
\includegraphics[width=12cm]{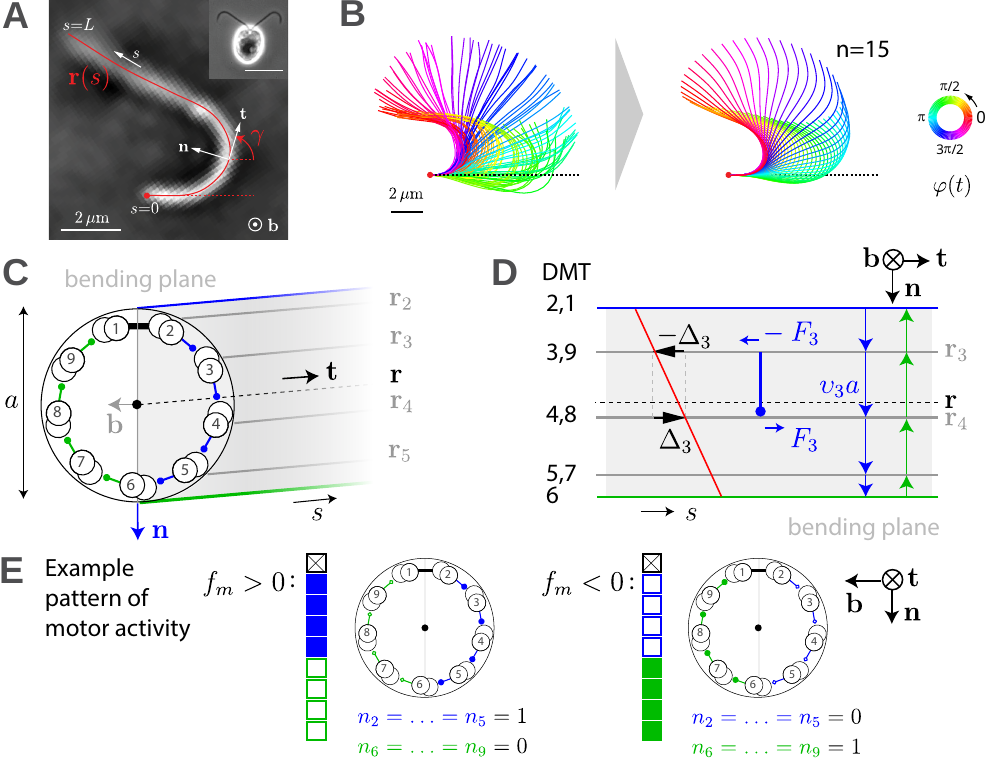}
\caption[]{
\textbf{Geometry of axonemal beating.}
\textbf{A.}
Micrograph of free-swimming reactivated \textit{Chlamydomonas} axoneme
with centerline $\rvec(s)$ parametrized by arclength $s$ (red line) and shear angle $\gamma(s)$.
The shear angle characterizes the direction of the local tangent $\tvec=\partial_s\rvec$
relative to the base at $s=0$ (red dot).
Inset: \textit{Chlamydomonas} cell, scale bar: $10\,\micron$.
\textbf{B.} 
Traced beat pattern of axonemal centerline $\rvec(s)$ 
aligned at basal end ($s=0$, red dot), color-coded by time $t$ (\textit{left}), and
phase-averaged beat pattern color-coded by oscillator phase $\varphi$ mod $2\pi$ (\textit{right}).
Beat frequency $f_0 = 68.7\pm 10.8\,\Hz$ (mean$\pm$SD, $n{=}15$ wildtype axonemes)~\cite{Sharma2024}.
\textbf{C.} 
Schematic of the axoneme with 9-fold symmetry of doublet microtubules (DMT, respective centerlines $\rvec_j$), 
connected by dynein molecular motors (blue/green). 
A sliding restriction between DMTs 1 and 2 (black) is supposed to define the bending plane (gray)~\cite{striegler_twisttorsion_2025}.
\textbf{D.} 
Motor-induced relative sliding between adjacent DMTs $j$ and $j+1$ with active motor force $F_j$ 
and local sliding displacement $\Delta_j$ causes bending of the axoneme in its bending plane.
If DMTs retain their distance and angular position relative to the axonemal centerline,
the sliding displacements are uniquely determined by the shear angle $\gamma(s)$, see Eq.~\eqref{eq:Delta_j}. 
Modified from \cite{Kotz2026}.
\textbf{E.} 
Example pattern of motor activity corresponding to maximal load sharing:
at any instance in time, all motors on one side of the axoneme are either fully active ($n_j=1$) or inactive ($n_j=0$).
Estimates on the force per active motor head derived below assume that all dynein motors are equally distributed among DMTs $2-9$, 
yet are independent of any assumption of the pattern of motor activity.
Axonemal cross-section viewed from the base. 
}
\label{fig1}
\end{figure*}

Molecular motors are protein machines that convert chemical energy into mechanical work~\cite{Alberts:book,Howard2002}. 
They transport cargo inside cells~\cite{Vale1985}, drive sarcomere contraction in muscle cells~\cite{Huxley1954}, and
generate mechanical tension in the acto-myosin cytoskeleton \cite{Murrell2015}.
Inside $\micron$-long motile cilia of eukaryotic cells, collections of $\sim 10^4$ dynein molecular motors drive regular bending waves at frequencies $10-100\,\Hz$ 
that propel microswimmers in a liquid or pump fluids in multicellular organisms~\cite{Gray1928,Lauga2009}.
Single-molecule experiments revealed typical maximal forces exerted by single motor domains in the range $1-5\,\pN$~\cite{Svoboda1993,Schnitzer1997}, 
including experiments on dynein motors isolated from cilia~\cite{sakakibara_inner-arm_1999,hirakawa_processive_2000,Kojima2002}.
However, probing force generation \textit{in situ} for physiological geometric arrangements of molecular motors and cytoskeletal filaments remains challenging~\cite{fujiwara_versatile_2023}.
The cytoskeletal core of motile cilia and eukaryotic flagella, the \textit{axoneme}, displays a stereotypic arrangement of dynein motors, 
which are regularly distributed along the entire length of 9 doublet microtubules (DMT);
the DMTs in turn are arranged in cylindrical fashion~\cite{bui_polarity_2012}, see Fig.~\ref{fig1}. 
The known number of dynein heads in the axoneme, 
together with an estimate of the active force needed to bend a beating axoneme, 
allows to infer a lower bound for the peak force generated per active motor head.

Already in 1958, Machin concluded that active motor forces must be generated along the entire length of the cilium~\cite{Machin1958}.
Subsequent work provided refined estimates of the profile of active motor forces $f_m(s)$~\cite{Brokaw1971,Hiramoto1978,Johnson1979}. 
Cryoelectron microscopy revealed the highly regular arrangement of different dynein subtypes and their total number~\cite{bui_polarity_2012}, 
together with indications for their state of activity~\cite{Lin2018,Howard2022}. 
Since the total number of dynein heads is precisely known, 
we can use this information to derive a lower bound 
for the force exerted by an individual dynein head in beating axonemes.
Our analysis indicates that the peak force per active dynein head exceeds a lower bound of $5\,\pN$, even when using the most conservative estimates.
This estimate exceeds the typical stall forces $1-5\,\pN$ of molecular motors measured in single-molecule experiments~\cite{sakakibara_inner-arm_1999,hirakawa_processive_2000,Kojima2002}.
Our estimate suggests that the dense motor collectives in a 96-nm repeat of the axoneme might be capable of force generation that is more efficient than those typically observed for individual motor heads. 
Alternatively, the axoneme might be much softer than previously assumed~\cite{Howard2002,Okuno1981,Gittes1993,Ishijima1994,Mickey1995,Xu2016}.

\paragraph*{The force balance of axonemal beating.} 
An extensive literature of mathematical models describes the motor-driven, regular bending waves of axonemes~
\cite{Hines.Blum1978,Lindemann2004,Brokaw2009,Camalet2000,Riedel2007,Sartori2016,Oriola2017,Geyer2022,Cass2023}.
Irrespective of an on-going discussion on the precise mechanisms of motor control~\cite{lindemann2010flagellar}, 
virtually all models are based on a local force balance between passive elastic forces and active motor forces, which we review now. 

In many important cases, cilia beat patterns are approximately planar, including the 
reactivated \textit{Chlamydomonas} axonemes analyzed here~\cite{Sharma2024}.
Up to translations and rotations, planar axonemal beat patterns are fully characterized by their shear angle profile, 
which measures the angle between the tangent $\tvec=\partial_s\rvec$ of the centerline $\rvec(s)$ at its arclength position $s$ and the tangent $\tvec(s=0)$ at the base, see Fig.~\ref{fig1}A.
The local curvature is then given by $\kappa=\partial_s \gamma$. At the length- and time-scales of cilia beating, inertia is negligible, 
so Newton's second law reduces to an instantaneous balance of force moments for each infinitesimal axonemal segment. 
This force balance reads for the standard planar sliding-filament description~\cite{Machin1958, Riedel2007, Cass2023, Kotz2026}
(comprising bending elasticity, sliding elasticity, hydrodynamic friction forces, internal sliding friction forces, and active motor forces)
\newcommand{\underbracetext}[3]{ \underbrace{ \strut #1 }_{\substack{\text{#2} \\ \text{#3}}} }
\begin{equation}
\underbracetext{ B\, \partial^2_s\, \gamma }{bending}{elasticity} 
\,-\, \underbracetext{ K a^2\, \gamma }{sliding}{elasticity} 
\,-\hspace{-3mm}  \underbracetext{ \mathcal{H} }{hydrodynamic}{friction} 
\hspace{-4mm}
\,-\, \underbracetext{ \mathcal{F} }{sliding}{friction}
\hspace{-1mm} \,+\,
\underbracetext{ a\,f_m}{motor}{force} = 0 \,,
\label{eq:force_balance}
\end{equation}
where $a\approx 200\,\nm$ denotes the axonemal diameter.
We discuss the different force contributions in sequential order. 
First, for a linearly elastic axoneme with bending stiffness $B$, the bending moment equals $B\kappa = B\partial_s\gamma$. 
Its spatial derivative $B\partial^2_s\gamma$ enters the balance of local torques acting on each infinitesimal axonemal segment~\cite{Cass2023}.
Second, for an axoneme where neighboring DMTs can slide relative to each other, 
a sliding elasticity contribution $K a^2 \gamma$ with sliding stiffness $K$ is expected due to the presence of elastic linkers connecting the DMTs. 
In fact, the so-called counterbend phenomenon, where a local deflection of the axoneme induces an opposite deflection more distally~\cite{Lindemann2005,Gadelha2013}, 
could be quantitatively described by this linear model, which further provided the estimate $K = 2000\,\pN / \micron^2$~\cite{Xu2016}.
Third, hydrodynamic friction force moments are computed as 
$\mathcal{H}(s,t)=-\n(s,t)\cdot\int_s^L\!ds'\,\f_\mathrm{hydro}(s',t)$, 
where $\f_\mathrm{hydro}(s',t)$ denotes the line density of hydrodynamic friction forces acting on an infinitesimal segment of the axoneme at arc-length position $s'$ and time $t$, 
and $\n$ the local normal vector perpendicular to $\tvec$.
The hydrodynamic friction forces $\f_\mathrm{hydro}$ can be computed using various approximation schemes if the shape dynamics $\gamma(s,t)$ is known~\cite{Lauga2009}. 
Fourth, for full generality, we further allow for an internal sliding friction term, 
for which the only assumption is that this force moment has the same sign as $\partial_t\gamma$. 
Indeed, experiments indicate the presence of internal friction opposing sliding~\cite{Klindt2016, Mondal2020, Pellicciotta2020}. 
Finally, $f_m(s,t)$ represents the active motor force generated by the collection of molecular motors.  
Competing mathematical models made different proposals for the dependence of $f_m$ on the local deformation of the axoneme. 
Here, we work backwards to estimate $f_m(s,t)$ from data using Eq.~\eqref{eq:force_balance}.

\begin{figure*}
\includegraphics[width=18cm]{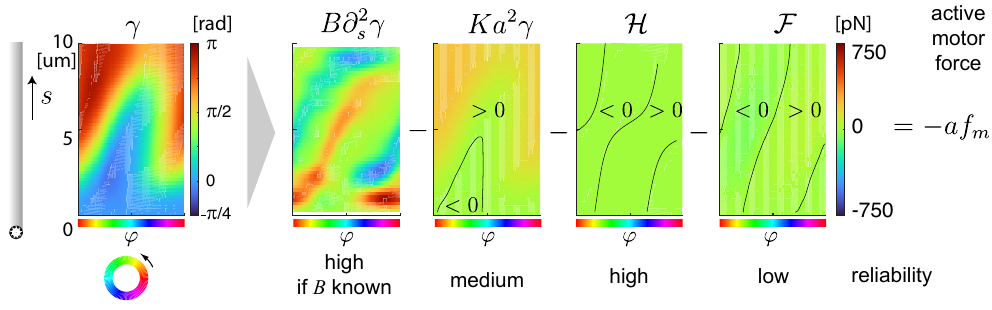}
\caption[]{
\textbf{Time-dependent passive forces inferred from measured beat patterns.}
Kymograph of shear angle $\gamma(s,t)$ from wildtype phase-averaged beat pattern from Fig.~\ref{fig1}B. 
From this data, we can estimate 
the bending force moment $B\partial_s^2\gamma$, sliding force moment $Ka^2\gamma$, 
hydrodynamic friction force moment $\mathcal{H}$, and sliding friction force moment $\mathcal{F}$, 
see Eq.~\eqref{eq:force_balance}. 
Parameters: $B=840\,\pN\,\micron^2$~\cite{Xu2016}, $K=2000\,\pN/\micron^2$~\cite{Xu2016}, 
$b=0.53\,\pN\,s/\micron$~\cite{Kotz2026}.
The reliability for estimation of these force moments is indicated.
}
\label{fig2}
\end{figure*}

The different force moment contributions in Eq.~\eqref{eq:force_balance}
are visualized as kymographs as functions of beat cycle phase $\varphi$ and arclength $s$ in Fig.~\ref{fig2}.
Unless stated otherwise, all waveform-based analyses use phase-averaged beat patterns of
free-swimming, reactivated \textit{Chlamydomonas} axonemes without motor extraction, 
at the maximal ATP concentration $\mathrm{[ATP]}=750\,\mu\mathrm{M}$, from Sharma et al.~\cite{Sharma2024}.
To minimize measurement noise, these phase-averaged beat patterns were further averaged across several axonemes,
both for wild-type cells (group-averaged beat pattern shown in Fig.~\ref{fig1}B) and for the ODA mutant lacking outer-arm dyneins.

\begin{figure*}
\includegraphics[width=14cm]{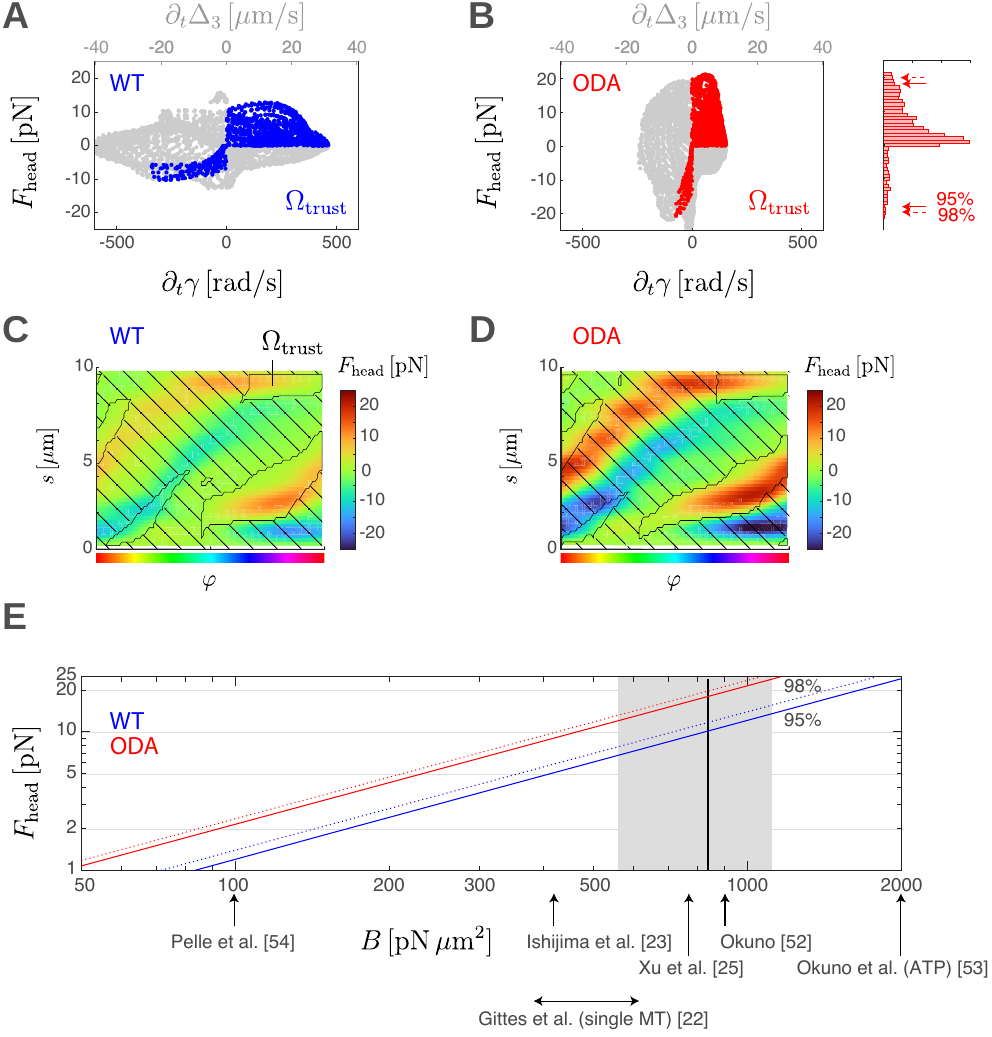}
\caption[]{
\textbf{Lower bound on motor force per active dynein head}
\textbf{A.} 
Scatter plot of local shear rate $\partial_t\gamma$ versus inferred signed lower bound $F_\text{head}$ 
on motor force per active dynein head
for wildtype beat patterns ($n=15$, data from~\cite{Sharma2024}),
using the conservative relation 
$F_\text{head} = -F_\text{passive,known}/(a\rhoDMT)$ from Eq.~\eqref{eq:Fhead}, 
where the known passive force moment $F_\text{passive,known}$ 
comprises both bending elasticity and hydrodynamic friction. 
The sign of $F_\text{head}$ provides the sign of the active motor force $a f_m$.
Data points highlighted in blue correspond to the region-of-trust $\Omega_\text{trust}$.
For these space-time points, force moments from sliding elasticity ($-Ka^2\gamma$) 
and sliding friction ($-\mathcal{F}$), which can be less reliably estimated, 
have the same sign as the known passive force moment,
hence would increase the lower bound on the force magnitude if included.
The horizontal axis on the top (gray) depicts sliding speed $\partial_t\Delta_3$ between DMTs 3 and 4 
(opposite for DMTs 8 and 9) corresponding to $\partial_t\gamma$ according to Eq.~\eqref{eq:Delta_j}.
Mean beat frequency $f_0 = 68.7\pm 10.8\,\Hz$ (mean$\pm$SD).
\textbf{B.} 
Analogous to panel A, but for the ODA mutant lacking outer-arm dyneins
(phase-averaged beat pattern, averaged over $n=31$ axonemes, data from ~\cite{Sharma2024}).
Sliding speeds are lower, reflecting the lower mean beat frequency $f_0 = 29.5\pm 5.8\,\Hz$ (mean$\pm$SD).
Inferred motor forces are higher, reflecting the reduced density of motor heads in the ODA mutant $(43\%)$ compared to wildtype.
The histogram on the right displays the marginal distribution of $F_\text{head}$, 
together with the 95th and 98th percentiles of its absolute value used for subsequent analysis. 
\textbf{C.} 
Data from panel A replotted as kymograph as a function of oscillator phase $\varphi$ and arc-length $s$
(region-of-trust $\Omega_\text{trust}$ shown as non-hatched regions). 
\textbf{D.} 
Analogous to panel C, but for the ODA mutant.
\textbf{E.} 
Summary of lower bounds on peak motor force per active dynein head for wildtype (blue) and ODA mutant (red)
from panels A and B, depicting 95th percentile (solid) and 98th percentile (dashed) of all data points from the region-of-trust as a function of the axonemal bending stiffness $B$.
The value $B=840\pm 280\,\pN\,\micron^2$ from \cite{Xu2016} used in panels A-D 
(black: mean, gray: SD), as well as other estimates from the literature, are indicated. 
All analyses are based on phase-averaged beat pattern from averaging beat patterns of individual axonemes
($n=15$ for wildtype, $n=31$ for ODA); 
for analyses of individual axonemes, see Fig.~\ref{fig_Fhead_individual}.
}
\label{fig3}
\end{figure*}

\section{A lower bound for motor forces}
Our approach is based on a set of conservative assumptions.

\paragraph*{Motor number.} 
The number of dynein heads in a \textit{Chlamydomonas} cilium of length $L = 10\,\micron$ has been determined as 
$N_\text{WT} = 17'500$ for wildtype cells and $N_\text{ODA} = 42.9\%\,N_\text{WT}$ for the ODA mutant lacking outer-arm dyneins~\cite{bui_polarity_2012,Sharma2024}.
We assume that all dynein motors contribute to the generation of axonemal bending waves. 
This is a conservative assumption, as likely some dyneins contribute to a static bend of the axoneme~\cite{Geyer2016}, or are otherwise inactive. 
Thus, by assuming the maximal possible number of active motor heads, our calculation provides a conservative lower bound on the force per active motor head. 
Using the dynein distribution proposed by Bui~et~al.~\cite{bui_polarity_2012}, 
the total motor-head density varies only weakly along the length of the axoneme and across DMTs
(coefficient of variation $<5\%$, see also Table~\ref{table:Bui}).
We therefore approximate the motor density $\rho$ as homogeneous along the axonemal length. 
Below, we will further make different assumptions on how motors are distributed among the 9 DMTs.

\paragraph*{Geometry of axonemal beating.} 
We assume that the axonemes analyzed here beat in a plane. 
This assumption is supported by previous 3D tracking of reactivated \textit{Chlamydomonas} axonemes, 
which showed that out-of-plane components are small~\cite{striegler_twisttorsion_2025}.
We further assume that relative sliding of doublet microtubules is restricted at the basal end of the axoneme,
i.e., $\Delta(s{=}0,t)=0$~\cite{Satir1968, Vernon2004}. 
Under this basal constraint, the shear angle $\gamma(s,t)$ is directly determined by the experimentally measured dynamics of the axonemal centerline $\rvec(s,t)$. 
The assumption of zero sliding at the base can be relaxed; 
as shown below, the inferred force balance is dominated by bending-elasticity, 
which is independent of a time-dependent basal sliding offset.

Finally, we assume that all doublet microtubules retain a constant distance and angular position relative to the virtual centerline of the axoneme. 
With this geometric constraint, the shear angle profile $\gamma(s,t)$ uniquely determines the relative sliding displacement $\Delta_j(s,t)$ 
between each pair of neighboring doublet microtubules $j$ and $j+1$ according to~\cite{Kotz2026}
\begin{equation}
\Delta_j(s,t) = \upsilon_j a \gamma(s,t) \quad,
\label{eq:Delta_j}
\end{equation}
where $\upsilon_j$ are signed geometric prefactors with $0\leq |\upsilon_j| \lesssim 0.34$
(with $\upsilon_1 = 0$ and $\upsilon_3 = -\upsilon_8 \approx 0.34$),
see SI appendix.

\paragraph*{Active forces.} 
The active motor force moment density $a f_m(s,t)$ in Eq.~\eqref{eq:force_balance} 
is given by the sum of all force moments generated by molecular motors per unit axonemal length. 
Let $\rho_j$ denote the line density of dynein motor heads on DMT $j$, 
$n_j$ the fraction of these motors that are currently producing force, 
and $F_j$ the average force per active motor head. 
Correspondingly, a segment of length $ds$ of DMT $j$ contains
$\rho_j n_j\,ds$, which together generate a total active force $\rho_j n_j F_j\,ds$.
Our aim is to find a lower bound $|F_\text{head}|$ such that 
$F_j\ge |F_\text{head}|$ for at least one $j$.
Using the geometric relation Eq.~\eqref{eq:Delta_j}, the active force moment density 
can be written as~\cite{Kotz2026}
\begin{equation}
a f_m = \sum_{j=1}^9 a \upsilon_j \rho_j n_j F_j \quad. 
\label{eq:fm}
\end{equation}
To obtain a conservative lower bound $F_\mathrm{head}$ for the force per active motor head, 
we make use of the known distribution and directionality of dynein motors.
Bui et al. showed that the density of dynein motor heads varies only weakly across DMTs~\cite{bui_polarity_2012}.
Specifically, with dynein-f counted as two-headed, all other IDAs and minor dyneins as single-headed, 
and four three-headed ODAs per DMT per 96-nm repeat, 
we estimate that the total motor-head density is only $\approx 5\%$ lower in the proximal region than in the central/distal region, 
and that the coefficient of variation across individual DMTs is $\lesssim 5\%$~\cite{Kotz2026}.
Furthermore, motor density was found to be $10{-}15\%$ higher on DMTs 2-9 than on DMT 1 
(the position of a sliding restriction termed the 1-2 bridge)~\cite{bui_polarity_2012}. 

As a conservative assumption, 
we approximate the total motor-head density as homogeneous along the length of the axoneme
and assign it symmetrically to the eight DMTs 2-9. 
Thus, each of these DMTs is assigned a constant motor-head density 
$\rho_j=\rhoDMT$ with $\rhoDMT = \rho/8 \approx 220\,\micron^{-1}$ for $j=2,\ldots,9$.
This density is slightly larger than even the highest local motor-head density reported by \cite{bui_polarity_2012}.

We further assume that all active dyneins generate force in the same direction along their respective DMTs,
such that $F_j\geq 0$, consistent with the common directionality of dynein motors.
With this minimal assumption, 
Eq.~\eqref{eq:fm} implies a conservative lower bound for the force per active motor head.
With $\rho_j=\rhoDMT$ for DMTs $j=2,\ldots,9$, Eq.~\eqref{eq:fm} simplifies to
$f_m = \rhoDMT \sum_{j=2}^9 \upsilon_j n_j F_j$.
The geometric factors obey
\begin{equation}
\sum_{j=2}^5 \upsilon_j = -\sum_{j=6}^9 \upsilon_j \approx 0.97 < 1 \quad.
\end{equation}
Hence, for arbitrary motor activities $0\leq n_j\leq1$,
\begin{equation}
\left|\sum_{j=2}^9 \upsilon_j n_j F_j\right|
< \max_j F_j \quad.
\end{equation}
Consequently, the force per active motor head must satisfy the lower bound
\begin{equation}
\frac{|f_m|}{\rho_\mathrm{DMT}} < \max_j F_j \quad.
\end{equation}
Thus, the bound on $F_j$ must hold for the motors on at least one DMT.
Allowing motors to generate forces of either sign would reduce the lower bound by a factor of 2.

\paragraph*{Passive forces.} 
We assume that the force-balance equation Eq.~\eqref{eq:force_balance} holds at every arc-length position $s$ along a beating axoneme and at every time point $t$. 
We re-arrange this force balance by separating the active motor force density $f_m$ from passive force contributions. 
The passive force contributions can be further separated into a part that can be computed directly from experimentally measured beat patterns, 
$F_\text{passive,known}$, and a part that can only be estimated with additional assumptions, $F_\text{passive,unknown}$. 
Thus,
\begin{equation}
a |f_m| = | F_\text{passive,known} + F_\text{passive,unknown} | \quad . 
\end{equation}

\paragraph*{Lower bound on active forces.}
The key point is that a lower bound on the active motor force does not require computing $F_\text{passive,unknown}$ explicitly. 
It is sufficient to know its sign relative to $F_\text{passive,known}$. 
For that, it is sufficient to assume that $\mathcal{F}$ is positive for $\partial_t\gamma>0$, and negative otherwise.
We therefore restrict our analysis to those space-time points $(s,t)\in\Omega_\text{trust}$ in a region-of-trust for which $F_\text{passive,known}$ and $F_\text{passive,unknown}$ have the same sign. 
Mathematically, 
$\Omega_\text{trust} =\{ (s,\varphi) : \sign(B\partial_s^2\gamma-\mathcal{H})=-\sign(\gamma)=-\sign(\partial_t\gamma) \}$.
For these points,
\begin{equation}
a |f_m| \geq
\left|F_\text{passive,known}\right|
\quad
\text{for } (s,t)\in\Omega_\text{trust}
\quad .
\label{eq:ineq_trust}
\end{equation}
Thus, the known passive force contribution alone provides a conservative lower bound on the magnitude of the active motor force at these special space-time points.
Specifically, from previous high-precision tracking of planar beat patterns with axonemal centerline $\rvec(s,t)$~\cite{Sharma2024}, 
we can accurately compute the bending moment contribution $B\partial_s^2\gamma$ using previous estimates of the bending stiffness $B$~\cite{Xu2016}.
We can likewise compute the hydrodynamic friction moment contribution $\mathcal{H}$ from the measured centerline dynamics using standard resistive force theory~\cite{Gray1955} as detailed in~\cite{Kotz2026}. 
Together, these terms define the known passive contribution
\begin{equation}
F_\text{passive,known} =
B\partial_s^2\gamma - \mathcal{H} \quad. 
\label{eq:Fpassiveknown}
\end{equation}
The remaining passive terms in Eq.~\eqref{eq:force_balance}, 
namely the sliding-elastic contribution $K a^2 \gamma$ and the internal friction contribution $\mathcal{F}$, 
could likewise be estimated by adopting a value for the sliding stiffness $K$~\cite{Xu2016}, 
or by making specific assumptions about the functional dependence of internal friction on the local sliding speed,
for example a linear friction law $\mathcal{F} = ba\, \partial_t\gamma$ as in~\cite{Kotz2026}. 
However, to remain maximally conservative and general, we combine these terms into $F_\text{passive,unknown}$ and use only their sign.
Combining everything, we arrive at the main result
\begin{equation}
\begin{split}
|F_\mathrm{head}| < F_j
\text{ with } F_\mathrm{head} = - \frac{F_\text{passive,known}}{a\, \rhoDMT} \\
\text{for at least one } j \text{ and } (s,t)\in\Omega_\mathrm{trust}\quad.
\label{eq:Fhead}
\end{split}
\end{equation}
Thus, the magnitude $|F_\mathrm{head}|$ provides a lower bound on the peak force per active motor head,
while the sign of $F_\mathrm{head}$ provides the sign of the active motor force $a f_m$.

Fig.~\ref{fig3} displays $F_\mathrm{head}$ according to Eq.~\eqref{eq:Fhead} 
for both wildtype (Fig.~\ref{fig3}A) and the ODA mutant lacking outer-arm dyneins (Fig.~\ref{fig3}B), 
highlighting space-time points $(s,t)\in\Omega_\text{trust}$ within the region-of-trust.
For both wildtype and ODA, this region-of-trust covers approximately $25\%$ of all space-time points.
The envelope of this point cloud can be regarded as a lower-bound of the average force-velocity relation of dynein heads in beating axonemes. 
We find maximal sliding speeds on the order of $30\,\micron/\s$, 
consistent with \textit{in-vitro} experiments~\cite{sakakibara_inner-arm_1999,hirakawa_processive_2000,Kojima2002}. 
However, the inferred force per active motor head reaches $10{-}15\,\pN$.
At least for the ODA mutant, these peak force estimates occur close to zero sliding speed, as expected.

As a robust measure of the largest forces during the beat, 
we report the 95th percentile $F_\text{head}^\mathrm{95th}$ of $|F_\text{head}|$ 
(using exclusively space-time points from the region-of-trust)
rather than its noise-sensitive maximum.
We find $F_\text{head}^\mathrm{95th}=10.2\,\pN$ for wildtype axonemes 
and $F_\text{head}^\mathrm{95th}=18.2\,\pN$ for the ODA mutant with its reduced motor density.
These inferred forces exceed typical single-molecule stall forces of $1{-}5\,\pN$~\cite{sakakibara_inner-arm_1999,hirakawa_processive_2000,Kojima2002}.

The lower bound on apparent motor force per active dynein head displays a traveling wave pattern
as function of beat cycle phase $\varphi$ and arclength $s$, see Fig.~\ref{fig3}C,D.
This is expected for a propagating wave of dynein activity~\cite{Lin2018}.

\paragraph*{Dependence on bending stiffness.}
We repeated our analysis for different values of the axonemal bending stiffness $B$, see Fig.~\ref{fig3}E.
We find that the lower bound depends approximately linearly on $B$, 
consistent with a dominance of bending elasticity in the force moment balance Eq.~\eqref{eq:force_balance}.
Inferred peak motor forces scale approximately linearly with $B$, 
and match the expected range of $1{-}5\,\pN$ for $B=100{-}300\,\pN\,\micron^2$.

\section{Joint bounds on motor force and sliding speed}
\begin{figure*}
\centering
\includegraphics[width=13.6cm]{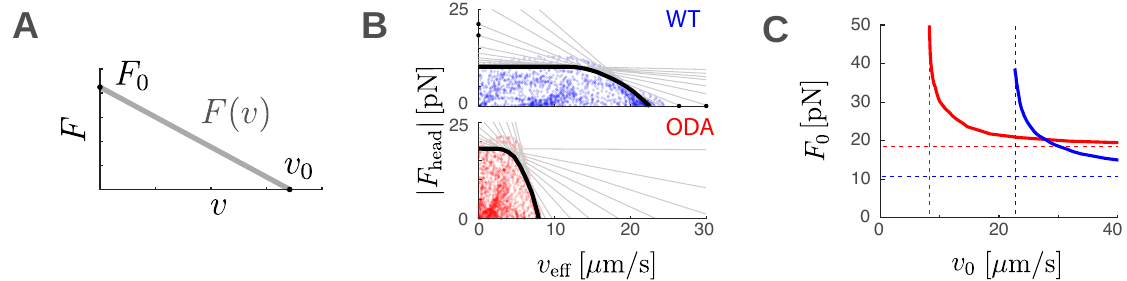}
\caption[]{
\textbf{Joint bounds on motor force and sliding speed.}
\textbf{A.}
Prototypical force-velocity curve given by Eq.~\eqref{eq:force_velocity}
with maximal force at zero velocity $F_0$ and velocity at zero load $v_0$ 
as assumed in models of axonemal beating~\cite{Riedel2007,Oriola2017,Cass2023,Kotz2026}.
\textbf{B.}
Absolute lower bound on force per active dynein head $F=|F_\mathrm{head}|$ versus 
effective motor sliding velocity $v_\eff\approx 0.26\, a|\partial_t\gamma|$ 
for wildtype and the ODA mutant,
replotting data from Fig.~\ref{fig3}AB using only data points in $\Omega_\mathrm{trust}$.
Additionally shown are families of linear relations of the form Eq.~\eqref{eq:force_velocity},
each with $95\%$ of the data points below it (gray), together with their pointwise lower envelope (black).
\textbf{C.}
Trade-off between $F_0$ and $v_0$ for the family of linear force-velocity relations from panel B
for wildtype (blue) and the ODA mutant (red). 
Dashed lines indicate asymptotes; the horizontal asymptotes equal $F_\mathrm{head}^\mathrm{95th}$ by construction.
}
\label{fig4}
\end{figure*}

Previous single-molecule experiments characterized force-velocity relations of individual dyneins \textit{in vitro}~\cite{hirakawa_processive_2000,Kojima2002}.
From our analysis, we can attempt to constrain a force-velocity relation 
in a setting where dyneins act collectively within a beating axoneme. 
Several mathematical models of axonemal beating had assumed a linear force-velocity relationship 
for active dynein motors~\cite{Riedel2007,Oriola2017,Cass2023,Kotz2026}, sketched in Fig.~\ref{fig4}A,
\begin{equation} 
F = F_0 ( 1 - v/v_0 ) \quad.
\label{eq:force_velocity}
\end{equation} 
Here, the force $F_0$ at zero velocity plays the role of a stall force, 
while $v_0$ corresponds to the maximal motor speed at zero load.

Let us assume Eq.~\eqref{eq:force_velocity} holds for all motors on DMTs 2 to 9,
with identical $F_0$ and $v_0$.
We further assume a maximal load-sharing scenario, in which all motors on one side of the axoneme are active,
while those on the opposite side are fully inactive, see Fig.~\ref{fig1}E.
With these assumptions, Eqs.~\eqref{eq:Delta_j} and~\eqref{eq:fm} imply
$|f_m| = \rhoDMT F_0 [ U_1 - U_2 a |\partial_t\gamma|/v_0 ]$,
where
$U_1=\sum_{j|n_j=1} |\upsilon_j|\approx 0.97$
and
$U_2=\sum_{j|n_j=1} \upsilon_j^2\approx 0.26$, 
provided $f_m$ and $\partial_t\gamma$ have the same sign, as satisfied for space-time points in $\Omega_\mathrm{trust}$.

Together with the lower bound on $|f_m|$ in $\Omega_\mathrm{trust}$ in Eq.~\eqref{eq:ineq_trust} and 
the definition of $F_\mathrm{head}$ in Eq.~\eqref{eq:Fhead}, 
this yields (using $U_1\approx 1$)
\begin{equation}
|F_\mathrm{head}| 
\le F_0(1-v_\eff/v_0)
\quad\text{with}\quad
v_\eff\approx 0.26\,a|\partial_t\gamma|\quad.
\label{eq:fm_gammadot}
\end{equation}
We therefore plot $|F_\mathrm{head}|$ versus $v_\eff$ in Fig.~\ref{fig4}B.
From these plots, we determine linear relations of the form Eq.~\eqref{eq:fm_gammadot}
such that $95\%$ of data points fall below each line.
Each line represents a possible combination of stall force $F_0$ and zero-load velocity $v_0$.
The pointwise lower envelope of the family of possible linear force-velocity relations provides 
a lower limit on the force-velocity relation of axonemal dyneins.
This envelope itself encloses $86.8\%$ and $87.1\%$ of all data points for wildtype and the ODA mutant, respectively.
From these lines, we obtain combinations of stall force $F_0$ and zero-load velocity $v_0$
that are compatible with $95\%$ of the inferred force bounds, see Fig.~\ref{fig4}C.
A larger maximal force $F_0$ permits a smaller zero-load velocity $v_0$, and vice versa.
Assuming a convex force-velocity relation instead of Eq.~\eqref{eq:force_velocity}, 
as suggested by~\cite{hirakawa_processive_2000},
would increase these joint bounds.

\section{Additional tests}

\paragraph*{Individual axonemes.}
We repeated our analysis using phase-averaged beat patterns of individual axonemes and find results similar to Fig.~\ref{fig3} for both wildtype and the ODA mutant, see Fig.~\ref{fig_Fhead_individual}.

\paragraph*{Influence of static mean shape.}
Axonemal bending waves can be represented as a superposition of a static mean shape and a dynamic traveling wave, 
both of which are supposedly generated by the activity of molecular motors~\cite{Geyer2016}.
For our analysis, we used the full beat pattern with both static and dynamic component.
We verified that results virtually do not change if only the dynamic component is used (even if we assume that all dyneins contribute exclusively to this dynamic component), see Fig.~\ref{fig:SI_fig3_only_dynamic}.
This robustness of our results can be intuitively understood from the fact that 
the force moment balance is dominated by bending elasticity and that the static component
approximately represents a circular arc of constant curvature~\cite{Geyer2016}, 
yet a constant offset in curvature does not contribute to the bending elasticity force moment.

\paragraph*{Idealized beat pattern.}
We repeated our analysis with an idealized beat pattern
\begin{equation}
\gamma(s,t)= \bar{\kappa}\, s + a(s)\cos\!\left[2\pi\!\left(f_0 t-s/\lambda\right)\right]\quad,
\end{equation}
where we assume a linear amplitude profile $a(s)=a_{\max}\,s/L$, 
a constant static mean curvature $\bar{\kappa}$, and 
a constant wave speed $f_0\lambda$ with wavelength $\lambda$~\cite{Friedrich2010}.
This idealized waveform yielded somewhat smaller force bounds than the measured beat pattern, 
but of the same order of magnitude, namely
$F_\mathrm{head}^\mathrm{95th} = 3.1\,\pN$
using parameters matching wild-type beat patterns, and
$F_\mathrm{head}^\mathrm{95th} = 6.3\,\pN$
using parameters matching ODA mutant beat patterns, see also Fig.~\ref{fig:SI_fig3_idealized}.
Parameters were $a_{\max}=1.53$, $\bar{\kappa}L=2.49$, $\lambda/L=1.98$ for wildtype, and
$a_{\max}=1.04$, $\bar{\kappa}L=2.07$, $\lambda/L=1.69$ for ODA.
The approximately three-fold lower force estimates for the idealized beat patterns compared to those
for the measured beat patterns can be attributed to the joint effects of 
neglecting higher harmonics (15\% reduction), 
approximating the amplitude profile $a(s)$ by a linear profile (21\%), 
assuming a constant curvature $\bar{\kappa}$ of the static mean shape (7\%), and
assuming a constant wave speed $\lambda f_0$ (23\%).
Because these analytically prescribed, idealized waveforms are smooth,
this test eliminates tracking artifacts that could otherwise be amplified by numerical differentiation.
The fact that the inferred forces remain of the same order of magnitude supports the robustness of the inferred force scale.

\paragraph*{Order-of-magnitude estimate.}
A simple rule-of-thumb estimate captures the magnitude of the inferred forces~\cite{Kotz2026}.
For a prototypical traveling-wave beat pattern
$\gamma = A \cos[2\pi(f_0 t - s/L)]$ with 
wavelength equal to cilia length $L\approx 10\,\micron$~\cite{Lee2025} and 
amplitude $A\approx 0.5\,\mathrm{rad}$~\cite{Sharma2024}, 
we estimate curvature gradients
$\partial_s\kappa = \partial_s^2\gamma = A(2\pi/L)^2 \approx 0.2\,\micron^{-2}$.
In the beat patterns analyzed here, local curvature gradients even exceed 
$\partial_s\kappa\sim 0.5\,\micron^{-2}$. 
Hence $B\partial_s\kappa \sim 500\,\pN$ for $B\sim 1000\,\pN\,\micron^2$.
If all dynein motor heads on one side of the axoneme are active at a given instant in time, 
with line density $\rho/2\sim 1000\,\micron^{-1} = 1\,\nm^{-1}$ and 
effective lever arm length $a/4 \approx 50\,\nm$, 
we conclude $F_\text{head} \sim B\partial_s\kappa\,/\,[( \rho/2 ) (a/4)] \sim 10\,\pN$.

\paragraph*{Effective lever arm length.}
For our analysis, we used the full projected distance between the centerlines of doublet microtubules 
as effective ``lever arm'' length of dynein motors; 
if this length is reduced by one microtubule diameter ($25\,\nm$), 
lower bounds on motor forces would increase by 60\%.
We further used a conservative assumption for the distribution of dynein motors, consistent with experiments~\cite{bui_polarity_2012}.
Note that even in an unrealistic scenario where all dynein motors are assigned to the DMTs with the largest effective ``lever arm'' length, i.e., DMTs 3 and 8,
estimated forces would decrease by only 26\%, and thus remain surprisingly high.

\paragraph*{Basal sliding.}
For our analysis, we assumed zero sliding displacement at the basal end,
i.e., $\Delta_0=\Delta(s{=}0,t)=0$.
This assumption can be relaxed.
For this, we consider a quasistatic model of linear basal compliance with effective spring constant $k_b$, 
such that the elastic energy of the axoneme reads
\begin{equation}
	E_\mathrm{elastic} =
	\frac{k_b}{2} \Delta_0^2 +
	\frac{K}{2} \int_0^L\!ds\,\Delta(s)^2 +
	\frac{B}{2} \int_0^L\!ds\, [\partial_s \gamma(s)]^2,
\end{equation}
where $\Delta_0(t)$ denotes the basal sliding and the sliding displacement now reads
$\Delta(s,t) = \Delta_0(t) + a\gamma(s,t)$.
For known $\gamma(s,t)$ and a given value of $k_b$,
minimizing $E_\mathrm{elastic}$ with respect to $\Delta_0$ gives
$\Delta_0/a = - KL/(KL+k_b)\langle\gamma\rangle_s$.
Repeating our analysis for $0\le KL/(KL+k_b)\le 1$ showed that the inferred peak forces change only modestly
($F_\mathrm{head}^\mathrm{95th}=10.2{-}12.1\,\pN$ for wild-type,
$F_\mathrm{head}^\mathrm{95th}=18.1{-}21.3\,\pN$ for ODA).
This can be understood intuitively:
a non-zero basal sliding displacement changes the force moment due to sliding elasticity, 
but not the force moment due to bending, which dominates the force balance.
The region-of-trust does change, but this has only a small effect on the inferred peak forces.

\paragraph*{Weakly non-planar beating.}
We repeated our analysis with the high-precision beat pattern from~\cite{striegler_twisttorsion_2025}
and obtained a similar estimate of $F^\mathrm{95th}_\mathrm{head} = 7.1\,\pN$. 
This beat pattern was obtained by averaging $n=17$ reactivated wild-type \textit{Chlamydomonas} axonemes.
As in Fig.~\ref{fig3}, this estimate corresponds to the 95th percentile, assuming $B=840\,\pN\,\micron^2$. 
For this analysis, we neglected the comparatively small out-of-plane component of the beat pattern
(SD of z-coordinate $\approx 0.12\,\micron$).
Striegler~et~al.\ quantified twist in reactivated \textit{Chlamydomonas} axonemes, suggesting small-amplitude twist waves with maximal twist
$\Omega_3 \approx 0.4{-}0.5\,\mathrm{rad}/\micron$~\cite{striegler_twisttorsion_2025}. 
Twist adds an equal contribution 
$\Delta_\mathrm{twist} = \sin(2\pi/9)(a/2)^2\,\Omega_3$ 
to each sliding displacement $\Delta_j$~\cite{Sartori2016}. 
We estimate a maximal twist-induced sliding displacement
$\Delta_\mathrm{twist} \sim 3\,\nm$,
which is substantially smaller than typical sliding displacements induced by bending waves,
$|a\upsilon_j\gamma| \sim 30\,\nm$.

Instead of neglecting the small out-of-plane component of this beat pattern,
we can consider bending within the local bending plane. 
We define a shear angle 
$\gamma_\mathrm{3D}(s,t) = \int_0^s \!ds'\, \kappa_\mathrm{3D}(s',t)$
that corresponds to the planar beat pattern obtained by mathematically untwisting the local bending plane, 
where $\kappa_\mathrm{3D}(s,t)$ denotes the signed curvature of the three-dimensional axoneme centerline
(its magnitude equals the usual curvature, while its sign indicates the local sense of bending~\cite{striegler_twisttorsion_2025}).
Using $\gamma_\mathrm{3D}$ for an analogous analysis gives $F^\mathrm{95th}_\mathrm{head} = 6.1\,\pN$, 
indicating the robustness of our analysis.
A calculation using a more stringent region-of-trust, which requires that the twist-induced correction $\Delta_\mathrm{twist}$ does not reverse the sign of any individual sliding displacement $\Delta_j$, gave similar force estimates, see SI text.

Generally, while beat patterns of \textit{Chlamydomonas} cilia can be weakly non-planar,
we do not expect that twist substantially affects our estimate.
Under the experimentally observed twist-torsion coupling,
the cross-section of the axoneme locally co-rotates with the bending plane~\cite{striegler_twisttorsion_2025}.
Thus, relation Eq.~\eqref{eq:Delta_j} for the bending-induced contribution to filament sliding
holds to linear order also for twisted axonemes.
More generally, bending and twist correspond to different azimuthal ``Fourier'' modes of interdoublet sliding:
to linear order, twist corresponds to the zeroth mode $\sum_j \Delta_j$,
while bending corresponds to an orthogonal bending mode $\sum_j\upsilon_j\Delta_j$~\cite{SartoriTwist2016,striegler_twisttorsion_2025}.
As a consequence, twist does not modify the bending-associated sliding elasticity to linear order~\cite{SartoriTwist2016}, see also SI text.

\paragraph*{Long-wavelength sperm flagella.}
We repeated our analysis also with a data set of free-swimming bull sperm from~\cite{Riedel2007}
(flagellar length $\approx 60\,\micron$) 
and obtained the substantially smaller estimate
$F^\mathrm{95th}_\mathrm{head} = 0.72\pm 0.14\,\pN$ 
(mean$\pm$SD of 95th percentiles for $n=7$ sperm cells, assuming $B=840\,\pN\,\micron^2$).
Hydrodynamic friction force moments were computed as in~\cite{Friedrich2010},
approximating the sperm head as a prolate spheroid with semi-major axis $5\,\micron$ 
and semi-minor axes $2.5\,\micron$, and using the hydrodynamic friction coefficients from~\cite{Friedrich2010} without temperature correction.
Without hydrodynamic friction, we obtained
$F^\mathrm{95th}_\mathrm{head} = 0.44\pm 0.23\,\pN$. 

The main reason for the difference in the motor forces inferred for \textit{Chlamydomonas} cilia and bull sperm 
is the almost 6-fold longer wavelength $\lambda$ of the bull sperm beat pattern.
For comparable angular amplitudes, the contribution of bending elasticity to Eq.~\eqref{eq:force_balance} scales as
$\partial_s^2\gamma \sim \lambda^{-2}$;
thus, a 6-fold increase in wavelength reduces this contribution by approximately 36-fold.
Hence, the high force estimate is reproduced in an independent \textit{Chlamydomonas} data set,
whereas applying the same analysis to the long-wavelength beat of bull sperm yields substantially smaller motor forces.

\section{Discussion}
\paragraph*{Summary.}
Here, we derived a lower bound for the peak force generated by individual dynein heads in beating axonemes 
by combining two independent pieces of information: 
the known number and arrangement of dynein motors inside axonemes~\cite{bui_polarity_2012}, and
previous estimates of the elastic properties of axonemes~\cite{Xu2016}.
This analysis revealed a discrepancy:
our lower bound for the active force per dynein head ($10{-}15\,\pN$) exceeds 
previous estimates of dynein stall forces from single-molecule experiments ($1{-}5\,\pN$)~\cite{sakakibara_inner-arm_1999,hirakawa_processive_2000,Kojima2002}.
It also exceeds estimates from partially disintegrated axonemes ($0.1\,\pN$)~\cite{fujiwara_versatile_2023}, as well as
estimates based on the force required to arrest flagellar beating ($5\,\pN$)~\cite{Schmitz2000}.
This discrepancy implies either that
(i) the bending stiffness of axonemes was previously overestimated, or that
(ii) collectives of molecular motors can generate forces more effectively in intact, beating axonemes.
In the following, we discuss both possibilities.

\newcommand{\BMT}{B_\text{MT}}
\paragraph*{Bending stiffness of the axoneme.}
Our analysis singled out bending elasticity as the main contributor to the force moment balance in beating axonemes (Fig.~\ref{fig2}). 
In contrast, hydrodynamic friction forces are negligible for free-swimming reactivated axonemes, 
consistent with previous work~\cite{Mondal2020,Cass2023,Kotz2026}.
Consequently, the estimated lower bound for the peak force $\Fhead$ per active dynein head
approximately scales with the bending stiffness $B$ of the axoneme (Fig.~\ref{fig3}E). 

In our estimate for the motor force, we used the recent estimate $B=840\pm 280\,\pN\,\micron^2$ 
for wild-type \textit{Chlamydomonas} axonemes rendered immotile by vanadate \cite{Xu2016}.
At least two factors complicate the measurement of $B$.
First, dyneins crosslink neighboring DMTs, thus resisting interdoublet sliding and increasing the apparent axonemal stiffness.
As an extreme example, for dyneins in rigor state (no ATP), axonemes appear more than ten-fold stiffer~\cite{Okuno1980, Okuno1979}.
Even for vanadate-treated axonemes, dyneins will be trapped in a weakly bound state
which likely increases the apparent axonemal stiffness~\cite{Xu2016}.
Second, axonemes are not elastic Euler beams. 
Already the force balance equation Eq.~\eqref{eq:force_balance} combines contributions from 
linear bending elasticity with bending stiffness $B$ and linear sliding elasticity with sliding stiffness $K$.
Some authors even argued for a nonlinear sliding stiffness, which would render estimation of $B$ even more challenging~\cite{Hines.Blum1978}.

Assuming Eq.~\eqref{eq:force_balance}, 
estimates of $B$ as reported in Xu~et~al.~\cite{Xu2016} require a three-step method as pioneered in~\cite{Pelle2009}:
(i) a measurement of an \textit{apparent} stiffness that combines contributions from bending and sliding stiffness,
e.g., by deflecting the axoneme with a calibrated microneedle~\cite{Okuno1980} or optical tweezers~\cite{Xu2016},
(ii) an estimate of the relative contribution of bending and sliding stiffness, i.e., the ratio $K/B$, and finally
(iii) a computational model that allows to infer $B$ and $K$ from the information obtained in steps (i) and (ii).
For step (ii), commonly the counterbend phenomenon is exploited~\cite{Pelle2009,Xu2016}, 
where an externally imposed bend of the axoneme causes a bend in the opposite direction more distally 
as a result of sliding elasticity~\cite{Lindemann2005,Gadelha2013}.
Pelle~et~al.~had used this strategy with measurements for step (i) from~\cite{Okuno1980}
to infer the estimate $B\sim 100\,\pN\,\micron^2$~\cite{Pelle2009}, 
substantially below the estimate of Xu~et~al.~\cite{Xu2016}.
Although the image data used by Pelle~et~al.\ for step (ii) might have been of lower quality than that used by Xu~et~al., 
both studies obtained similar values for $K/B$ ($1.5\,\micron^{-4}$~\cite{Pelle2009}, $2.4\,\micron^{-4}$~\cite{Xu2016}).
Thus, the cause for the different estimates for $B$ could rather be the force measurements used for step (i).
We expect that immobilization by vanadate, as used by Xu~et~al.~\cite{Xu2016}, predominantly affects the sliding stiffness $K$, rather than the bending stiffness $B$, because residual dynein interactions can act as crosslinks between DMTs and contribute to resistance against interdoublet sliding. This expectation is consistent with their analysis of axonemal mutants: the inferred shear stiffness differed significantly between wildtype and mutants lacking the N-DRC or dynein arms, whereas the inferred intrinsic bending stiffness was not significantly different and remained within the range $777{-}1011\,\pN\,\micron^2$.
This further supports our use of the same value of $B$ for wild-type and ODA-mutant axonemes.
 
Independent estimates of interdoublet shear elasticity were obtained from thermal fluctuations,
$K\sim 75\,\pN\,\micron^{-2}$~\cite{brokaw_flagellar_1990} using data from~\cite{Kamimura1989},
and from direct longitudinal-shear measurements,
$K\sim 2000\pm 800\,\pN\,\micron^{-2}$~\cite{minoura_direct_1999}.

An estimate $B\approx 1700\,\pN\,\micron^2$ for the bending stiffness of bull sperm flagella
was obtained by fitting a mathematical model of flagellar beating, 
with the force scale set by an estimated hydrodynamic friction coefficient~\cite{Riedel2007}.
Their fitted model parameters were consistent 
with dynein force estimates from single-molecule experiments~\cite{sakakibara_inner-arm_1999}, 
in agreement with our re-analysis of their waveform data.
This difference from the force estimates for \textit{Chlamydomonas} cilia can be attributed to the substantially longer wavelength of the bull sperm beat pattern.

For completeness, we quote the historical estimate $60\,\pN\,\micron^2$ for the apparent stiffness (flexural rigidity) of sea urchin sperm flagella~\cite{Rikmenspoel1966}. 
Similarly, other earlier measurements of the apparent stiffness reported 
$900\,\pN\,\micron^2$ for sea urchin sperm flagella in the presence of vanadate~\cite{Okuno1980}, 
$1900\,\pN\,\micron^2$ for echinoderm sperm flagella at intermediate ATP levels~\cite{Okuno1979}, and
$420\,\pN\,\micron^2$ for bending of echinoderm sperm (in the beat plane)~\cite{Ishijima1994}.
Magnetic-tweezer experiments of intact human airway cilia yielded an apparent flexural rigidity of $620\pm160\,\pN\,\micron^2$~\cite{hill_force_2010}.

Previous measurements on primary cilia (whose axonemal architecture differs from that of motile 9+2 cilia)
reported flexural rigidities in the range $10{-}50\,\pN\,\micron^2$~\cite{Schwartz1997,Battle2015,Resnick2016},
although basal compliance may confound such estimates~\cite{Young2012,Battle2015}.
Notably higher values of $210{-}310\,\pN\,\micron^2$~\cite{Downs2014}
and $300{-}450\,\pN\,\micron^2$~\cite{Katoh2023} have also been reported.

\paragraph*{Bending stiffness of single microtubules.}
The bending stiffness $\BMT$ of single microtubules provides an independent structural estimate for the bending stiffness of axonemes.
For an axoneme with 9 doublet microtubules and 2 central-pair singlet microtubules, 
the intrinsic bending stiffness is expected to scale as
\begin{equation}
B \approx 9 B_\text{DMT} + 2\BMT \quad .
\end{equation}
Using $\BMT = 21.5\,\pN\,\micron^2$ for single taxol-stabilized microtubules, corresponding to a persistence length of $\BMT/k_BT\approx 5\,\mathrm{mm}$~\cite{Gittes1993}, 
and assuming that bending stiffness scales with protofilament number,
we estimate
$B_\text{DMT} \approx (23/13)\BMT$, 
which gives
$B \approx 385\,\pN\,\micron^2$.
A geometric estimate based on the second moment of area of a doublet microtubule,
$B_\text{DMT} \approx 3\BMT$,
instead gives
$B \approx 620\,\pN\,\micron^2$.
Additional structures within the axoneme may increase this estimate for $B$ further,
while cross-links between DMTs are expected to contribute primarily to the sliding stiffness $K$.
However, there remains considerable uncertainty regarding the bending stiffness of single microtubules~\cite{Gittes1993,Mickey1995,Felgner1996,Pampaloni2006,Memet2018}, 
with apparent stiffness depending on microtubule length and deformation amplitude~\cite{Pampaloni2006,Memet2018}. 
For long, weakly bent microtubules, which may provide the most relevant comparison for bending of DMTs inside axonemes, 
values in the range $\BMT\sim 10{-}30\,\pN\,\micron^2$ appear representative~\cite{Gittes1993,Mickey1995,Memet2018}.

\newcommand{\fJ}{\mathrm{fJ}}
\paragraph*{Implications for energetic cost of axonemal beating.}
A passive elastic deformation of the axoneme stores elastic energy, 
which we assume cannot be efficiently recovered for active work when the axoneme unbends, and is instead largely dissipated.
Thus, by computing changes in this elastic energy during a beat cycle, 
one can infer a lower bound for the energy required for axonemal beating.
A previous calculation (adjusted to the value of $B$ from \cite{Xu2016}) gives 
$0.25\,\fJ$ per beat cycle for reactivated wildtype axonemes, 
corresponding to the equivalent of $0.2$ ATP molecules per dynein head per beat cycle
(if $100\,\mathrm{zJ}$ of chemical energy are released upon hydrolysis of a single ATP molecule~\cite{Howard2002}).
More indirectly, when fitting computational models of axonemal beating, the assumed value of $B$ sets an energy-scale that strongly affects fit parameters and computed rates of energy dissipation. 
Previous work by the authors (assuming $B=840\,\pN\,\micron^2$) gave an estimate for the dissipation rate
corresponding to an average of 3 ATP molecules hydrolyzed per dynein head per beat cycle~\cite{Kotz2026}.
Thus, the value $B$ of the axonemal bending stiffness not only has implications for the estimated motor force $\Fhead$ per active dynein head, 
but may also give a hint whether a dynein head hydrolyzes one or several ATP molecules per beat cycle.
This distinction is important because hydrolysis of several ATP molecules would imply persistent motor activity, 
reminiscent of processivity of individual motors.
While axonemal dyneins are generally considered non-processive, processive behavior was reported for selected dyneins
~\cite{sakakibara_inner-arm_1999,hirakawa_processive_2000}.
Previous direct measurements corroborated the order of magnitude of about one ATP molecule hydrolyzed per dynein head per beat cycle~\cite{Brokaw1967,Chen2015}.

\paragraph*{Collective force generation.}
Structural evidence suggests close interactions between dynein motors~\cite{Walton2023}, 
raising the intriguing possibility that individual dynein heads might not perform their power-strokes independently from each other. 
Close interactions between motors may enable motor activity to propagate along the axoneme as an ``excitation wave'', 
as assumed in some unorthodox models of axonemal beating~\cite{Murase1986,Sharma2024}.
Recent theoretical studies have likewise considered interactions between neighboring motors and their consequences for collective motor dynamics and beat precision~\cite{Costantini2024,Fanelli2026}.
However, it is not yet clear how such motor coordination could increase the force generated per active dynein head.

\paragraph*{Outlook.}
Reliable estimates of forces generated by molecular motors, as well as of material properties of cytoskeletal filaments such as their bending stiffness, 
are essential for a quantitative molecular interpretation of cell mechanics. 
Here, we harnessed an established mechanical description of axonemal beating, without committing to a specific hypothesis for motor control, 
to derive a lower bound on the apparent force generated per active dynein head. 
We report this lower bound as a function of the axonemal bending stiffness and find that, for previous estimates of this stiffness~\cite{Ishijima1994,Xu2016}, 
it exceeds the typical stall forces reported for isolated molecular motors~\cite{sakakibara_inner-arm_1999,hirakawa_processive_2000,Kojima2002}.

This apparent discrepancy suggests at least three possible interpretations, which should be explored in future work:
First, the bending stiffness of axonemes may have been overestimated. 
Since reported axonemal bending stiffnesses are consistent with estimates based on the bending stiffness of single microtubules, 
such a revision may also require a re-evaluation of microtubule stiffness estimates. 
Second, dense and tightly integrated motor units, such as the \mbox{96-nm} repeats comprising 168 dynein motor heads in wildtype \textit{Chlamydomonas} axonemes~\cite{bui_polarity_2012,Sharma2024}, 
may collectively generate forces that exceed expectations based on isolated motors. 
To resolve this discrepancy, it may also be important to investigate whether the widely accepted force balance equation, Eq.~\eqref{eq:force_balance}, 
despite its successful use in numerous modeling studies, might still miss relevant contributions.

\begin{acknowledgments}
BMF was supported by the Deutsche Forschungsgemeinschaft (DFG, German Research Foundation) under Germany's Excellence Strategy - EXC-2068-390729961, as well as through a Heisenberg grant (421143374).
Maximilian Kotz acknowledges support from the Studienstiftung des Deutschen Volkes. 
\end{acknowledgments}
\paragraph*{Declaration of interests.}
The authors declare no competing interests.


\bibliography{reaction_diffusion_flagella}

\clearpage
\onecolumngrid 

\makeatletter
\def\@oddfoot{\footnotesize Kotz \textit{ et al.}\ |\ Supplemental Material \hfill \,}
\def\@evenfoot{\@oddfoot}
\makeatother

\setcounter{section}{0}
\renewcommand{\theequation}{S\arabic{equation}}    
\setcounter{equation}{0}  
\renewcommand{\thefigure}{S\arabic{figure}}    
\setcounter{figure}{0}  
\renewcommand{\thetable}{S\arabic{table}}    
\setcounter{table}{0}  
\renewcommand{\thepage}{S\arabic{page}}    
\setcounter{page}{1}  
\nocite{}

\makeatletter
\def\@seccntformat#1{\csname the#1\endcsname.\quad}
\def\thesection{\Roman{section}}
\setcounter{secnumdepth}{2}
\makeatother

\begin{center}
\LARGE Supplemental Material for\\[2mm]
\Large 
Axonemal bending stiffness of \textit{Chlamydomonas} cilia implies\\ single-motor forces above $5\,\pN$
\\[2mm]
\large
Maximilian Kotz, Veikko F. Geyer, Benjamin M. Friedrich\\[1mm]
\end{center}

This PDF file includes: 
\begin{enumerate}
	\item Supplementary Figures S1-S6
	\item Supplementary Table S1
\end{enumerate}

\section{Appendix: Computation of hydrodynamic friction forces}
\begin{figure}[b]
\includegraphics[width=9cm]{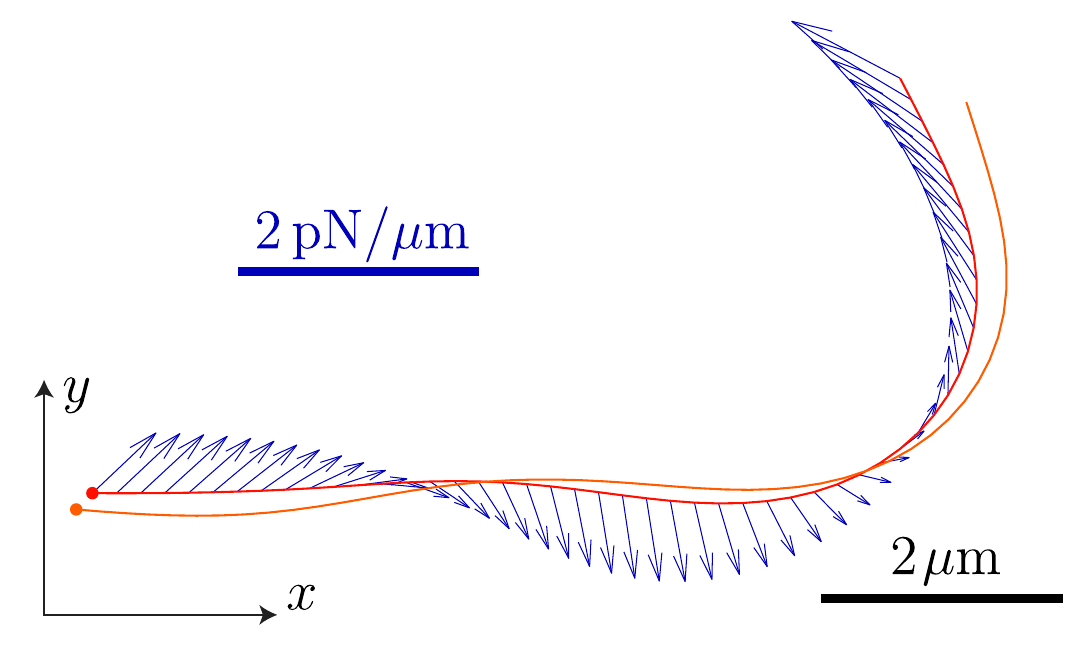}
\caption[]{
\textbf{Computed hydrodynamic friction forces.}
Hydrodynamic friction forces computed using resistive force theory according to Eq.~\eqref{eq:RFT} 
for the phase-averaged beat pattern of free-swimming reactivated \textit{Chlamydomonas} axonemes
from \cite{Sharma2024} shown in Fig.~\ref{fig1}B(right).
Shown are axonemal centerline $\rvec(s,t)$ at $t=0$ (red, corresponding to $\varphi=0$, basal end marked by red dot), and at $t=0.05\,T$ with mean beat period $T=f_0^{-1} \approx 14.6\,\mathrm{ms}$ (orange, corresponding to $\varphi=0.05\cdot 2\pi$, basal end marked by dot, rigid body motion determined from hydrodynamic computation), 
as well as computed line density of hydrodynamic friction forces $\f_\mathrm{hydro}(s)$ at $t=0$.
Scale bar indicates force magnitude: $2\,\pN/\micron$.
}
\label{fig_RFT}
\end{figure}

We use resistive-force theory to compute the line density of hydrodynamic friction forces $\f_\mathrm{hydro}(s)$ with units $\pN/\micron$ acting along the length of the cilium as~\cite{Gray1955}
\begin{equation}
\f_\mathrm{hydro}(s,t) 
= - \xi_\parallel\,(\mathbf{v}\cdot\tvec)\,\tvec - \xi_\perp\,(\mathbf{v}\cdot\n)\,\n \quad, 
\label{eq:RFT}
\end{equation}
where $\tvec$ and $\n$ denote the local tangent and normal vector of $\rvec(s)$, respectively, 
and $\mathbf{v}=\partial_t\rvec$ denotes the local velocity with respect to the laboratory frame.
For visualization, see Fig.~\ref{fig_RFT}.
For the hydrodynamic friction coefficients, $\xi_\parallel$ and $\xi_\perp$, we use the values
$\xi_\parallel=0.99\,\fN\,\s\,\micron^{-2}$, $\xi_\perp=1.81\,\xi_\parallel$
previously determined by a fit to data from bull sperm flagella~\cite{Friedrich2010}; 
these values are adjusted for temperature-dependent viscosity
by multiplication with a constant correction factor $1/0.7$
to account for the increased viscosity of water at the temperature
$T=20^\circ\mathrm{C}$ of the experiments in \cite{Sharma2024}, 
compared to the temperature $T=36^\circ\mathrm{C}$ of the experiments in \cite{Riedel2007,Friedrich2010}.
Further details of the hydrodynamic computations can be found in \cite{Kotz2026}.

\section{Appendix: Phase-averaged beat patterns}
We computed phase-averaged beat patterns for free-swimming, reactivated \textit{Chlamydomonas} axonemes 
using data from~\cite{Sharma2024} for both wildtype and ODA mutant axonemes without motor extraction.
For the analysis, shear angle profiles $\gamma(s,t)$ were computed from $(x,y)$ coordinates of traced axoneme centerlines as described in~\cite{Kotz2026}.
To average out fluctuations in the axonemal beat, 
we used a phase-averaging method described in~\cite{Werner2014}.
This method applies principal component analysis (PCA) to project the shear angle data 
on a low-dimensional shape space spanned by the principal shape modes.
First, the shear angle $\gamma(s,t)$ sampled at equidistant arc-length positions $s$
provides a feature vector in a high-dimensional feature space for each time point $t$.
PCA is applied to this set of feature vectors.
The PCA projection reveals a noisy limit cycle, 
reflecting the periodic nature of the axonemal beat. 
For the present data, the projection to the two principal shape modes retains more than $95\%$ of the total variance of $\gamma(s,t)$ for most individual beat patterns. 
A smooth, phase-averaged limit cycle was then fitted to this noisy limit cycle, 
using a polar coordinate representation of closed orbits, 
with radius $r(\overline{\varphi})$ as function of polar angle $\overline{\varphi}$ 
represented by a low-order Fourier series. 
The polar angle $\overline{\varphi}$ represents a so-called proto-phase,
which was mapped to a proper phase $\varphi(\overline{\varphi})$ using the method of~\cite{Kralemann_phase},
such that the mean phase speed conditioned on phase is constant,
$\langle d\varphi/dt \mid \varphi \rangle=\omega_0$,
where $\omega_0=2\pi\,f_0$ equals the angular frequency.
Importantly, each point in the two-dimensional PCA shape space has a unique reconstruction 
within the subspace spanned by the two principal shape modes; hence, the noise-averaged limit cycle in shape space naturally lifts to a phase-averaged beat pattern in the feature space of the shear angle.

\begin{figure}
\includegraphics[width=9cm]{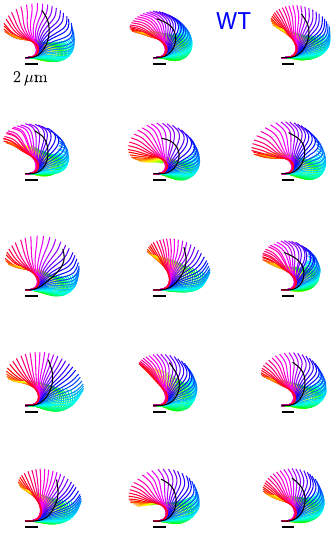}
\caption[]{
\textbf{Phase-averaged beat pattern of individual wildtype axonemes.}
Phase-averaged beat patterns of free-swimming reactivated \textit{Chlamydomonas} axonemes
from \cite{Sharma2024} used in this study, 
color-coded by oscillator phase $\varphi$; 
time-averaged mean shape shown in black. 
Mean beat frequency $f_0 = 68.7\pm 10.8\,\Hz$ (mean$\pm$SD, $n=15$).
Scale bar: $2\,\micron$.
}
\label{fig_beat_pattern_WT}
\end{figure}

\begin{figure}
\includegraphics[width=9cm]{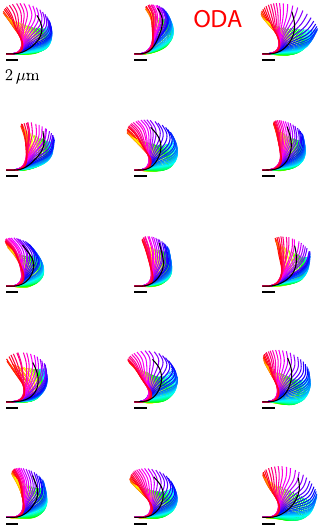}
\caption[]{
\textbf{Phase-averaged beat pattern of individual ODA axonemes.}
Phase-averaged beat patterns of free-swimming reactivated \textit{Chlamydomonas} axonemes
from \cite{Sharma2024} for the ODA mutant lacking outer-arm dyneins used in this study,
color-coded by oscillator phase $\varphi$.
Only 15 axonemes (random selection) out of a total of $n=31$ axonemes are shown;
time-averaged mean shape shown in black.
Mean beat frequency $f_0 = 29.5\pm 5.8\,\Hz$ (mean$\pm$SD, $n=31$).
Scale bar: $2\,\micron$.
}
\label{fig_beat_pattern_ODA}
\end{figure}

Phase-averaged beat patterns for free-swimming, reactivated \textit{Chlamydomonas} axonemes 
for wildtype and ODA mutant are shown in Fig.~\ref{fig_beat_pattern_WT} and Fig.~\ref{fig_beat_pattern_ODA}, respectively.
For the analysis in the main text, we additionally averaged the phase-averaged beat patterns of individual axonemes of a given condition, using their shear angle representation. 
For this group average, the proper phases $\varphi$ of the individual beat patterns were aligned to a common phase origin.
For this purpose, the shear-angle profiles were fitted by their first harmonic,
$\gamma(s,\varphi)=\bar{\gamma}(s)+a(s)\cos\left[\varphi-\Phi_0-2\pi s/\lambda\right]$.
The phase offset $\Phi_0$ obtained from this representation was used to determine the relative phase shift
between individual axonemes and to align all beat patterns to a common phase origin before averaging.

\section{Appendix: other beat patterns}
\begin{figure}
\includegraphics[width=9cm]{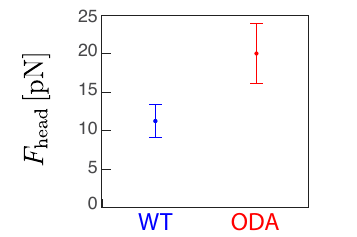}
\caption[]{
\textbf{Lower bound on motor force using beat patterns from individual axonemes.}
As a control, we repeated the analysis from Fig.~\ref{fig3} for beat patterns from individual axonemes, 
which provided a lower bound of the motor force per active dynein head for each axoneme
(computed as 95th percentile from analogously defined regions-of-trust).
Shown are mean$\pm$SD of these lower bounds for reactivated \textit{Chlamydomonas} axonemes for wildtype (blue, $n=15$) and ODA mutant (red, $n=31$).
Parameters: as in Fig.~\ref{fig3}, i.e., $B=840\,\pN\,\micron^2$~\cite{Xu2016}.
}
\label{fig_Fhead_individual}
\end{figure}

\begin{figure*}
\includegraphics[width=12cm]{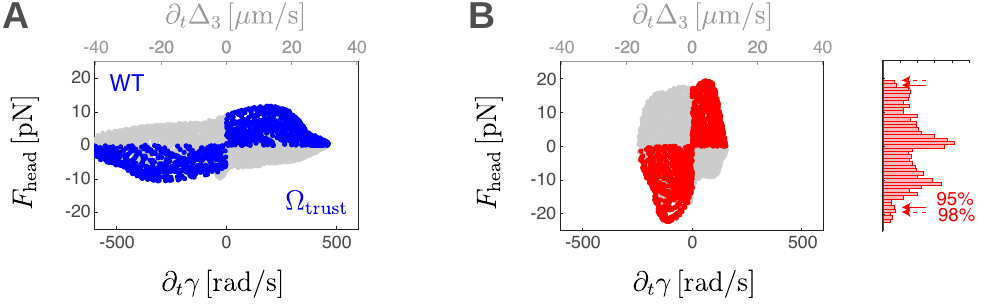}
\caption[]{
\textbf{Lower bound on motor force per dynein head, using only dynamic component of beat}
\textbf{A.} 
Analogous to Fig.~\ref{fig3}, 
scatter plot of local shear rate $\partial_t\gamma_D$
yet now for the dynamic component $\gamma_D(s,t)=\gamma(s,t)-\langle\gamma(s,t)\rangle_t$,
versus inferred signed lower bound $F_\text{head}$ 
on motor force per active dynein head for wildtype beat patterns ($n=15$, data from~\cite{Sharma2024})
using Eq.~\eqref{eq:Fhead}.
Data points highlighted in blue correspond to the region-of-trust $\Omega_\text{trust}$.
The horizontal axis on the top (gray) depicts sliding speed $\partial_t\Delta_3$ between DMTs 3 and 4 
corresponding to $\partial_t\gamma$ according to Eq.~\eqref{eq:Delta_j}.
$F_\mathrm{head}^\mathrm{95th}=9.6\,\pN$. 
\textbf{B.} 
Analogous to panel A again using $\gamma_D$, but for the ODA mutant lacking outer-arm dyneins,
(phase-averaged beat pattern, averaged over $n=31$ axonemes, data from ~\cite{Sharma2024}).
The histogram on the right displays the marginal distribution of $F_\text{head}$, 
together with the 95th and 98th percentiles.
$F_\mathrm{head}^\mathrm{95th}=18.0\,\pN$. 
}
\label{fig:SI_fig3_only_dynamic}
\end{figure*}

\begin{figure*}
\includegraphics[width=12cm]{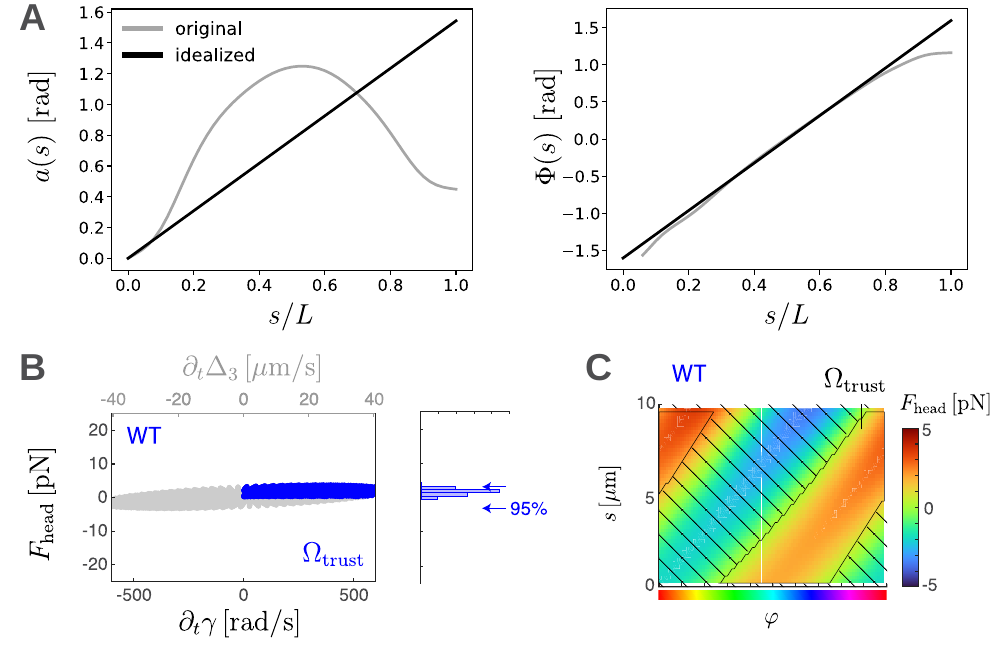}
\caption[]{
\textbf{Force inference for idealized beat pattern.}
\textbf{A.}
We repeated the analysis of Fig.~\ref{fig3} for idealized beat patterns. 
Generally, the principal Fourier model of the shear angle can be written as a wave
$\gamma(s,t) = a(s)\,\cos[\omega_0 t - \Phi(s)]$ 
with amplitude profile $a(s)$ and phase profile $\Phi(s)$~\cite{Kotz2026}.
We replaced the amplitude and phase profile of the phase-averaged wildtype beat pattern used in Fig.~\ref{fig3}A (gray) 
by linear approximations 
$a(s) = a_\mathrm{max}\,s/L$, $\Phi(s) = \Phi_0 + 2\pi s/\lambda$
(black) such that the amplitude $A$ (defined in terms of Fourier power~\cite{Kotz2026}), and
the slope $\partial_s\Phi(s)$ of the phase profile (determined by linear regression)
match for the phase-averaged and the idealized beat pattern.
\textbf{B.}
Scatter plot of local shear rate $\partial_t\gamma$
versus inferred signed lower bound $F_\text{head}$ 
on motor force per active dynein head using Eq.~\eqref{eq:Fhead}
for the idealized beat pattern, 
analogous to Fig.~\ref{fig3}A. 
Data points highlighted in blue correspond to the region-of-trust $\Omega_\text{trust}$.
The horizontal axis on the top (gray) depicts sliding speed $\partial_t\Delta_3$ between DMTs 3 and 4 
corresponding to $\partial_t\gamma$ according to Eq.~\eqref{eq:Delta_j}.
The histogram on the right displays the marginal distribution of $F_\text{head}$, 
together with the 95th percentile for $|F_\mathrm{head}|$,
$F_\mathrm{head}^\mathrm{95th}=3.1\,\pN$. 
Parameters: $A=0.88$, corresponding to $a_\mathrm{max}=1.53$, 
$\bar{\kappa}L=2.49$, and $\lambda/L=1.98$.
\textbf{C.} 
Data from panel B replotted as kymograph as function of oscillator phase $\varphi$ and arc-length $s$
(region-of-trust $\Omega_\text{trust}$ shown as non-hatched regions). 
Note the different scale of the color bar.
}
\label{fig:SI_fig3_idealized}
\end{figure*}

For the analysis in the main text, we used phase-averaged beat patterns that were additionally averaged over all individual axonemes of a given condition (wildtype visualized in Fig.~\ref{fig1}B).
As a control, we verified that the estimated lower bounds for the apparent peak motor force per active dynein head reported in Fig.~\ref{fig3} 
agree with the mean of analogously computed lower bounds computed for individual axonemes, see Fig.~\ref{fig_Fhead_individual}.
Using the phase-averaged, high-precision wildtype beat pattern, again for free-swimming reactivated \textit{Chlamydomonas} axonemes, from \cite{striegler_twisttorsion_2025}
provides a similar lower bound of $7.1\,\pN$ for the peak motor force per active dynein head (95th percentile).
This bound was computed using a projection of the weakly non-planar beat pattern on the $xy$-plane,
using the shear angle computed by integration of the signed curvature of the full three-dimensional waveform for the analysis gave a bound of $6.1\,\pN$.

\section{Appendix: Sliding displacements between doublet microtubules}
Specifically, we denote the centerlines of the 9 DMTs by $\rvec_j(s,t)$, $j=1,\ldots,9$ and the virtual centerline of the axoneme by $\rvec(s,t)$. 
The centerline $\rvec(s,t)$ is assumed to be planar with zero torsion and zero twist. 
Correspondingly, a canonical material frame is given by the Frenet-Serret frame of the centerline $\rvec$ consisting of a local tangent vector $\tvec=\partial_s\rvec$ 
and normal vector $\n = \bvec\times\tvec$, where $\bvec$ corresponds to the constant binormal vector perpendicular to the bending plane. 
The centerline of DMT $j$ is thus given by
\begin{equation}
\rvec_{j+6} = \rvec + \frac{a}{2} \left[ 
\cos\left(\frac{2\pi\,j}{9}\right) \n + \sin\left(\frac{2\pi\,j}{9}\right) \bvec 
\right] \quad 
\end{equation}
with indices understood modulo 9.
The sliding displacement $\Delta_j$ between DMT $j$ and DMT $j+1$ is given by the difference in their arc-lengths
\begin{equation}
\Delta_j(s) = \int_0^s \!ds'\, \left[ |\partial_{s'}\rvec_j(s')| - |\partial_{s'}\rvec_{j+1}(s')| \right]
= \upsilon_j a \gamma(s) \quad, 
\label{eq:Delta_j_def}
\end{equation}
where we introduced geometric factors 
\begin{equation}
\upsilon_j = \frac{1}{2} \left[ \cos\left(\frac{2\pi}{9} (j-5) \right) - \cos\left(\frac{2\pi}{9} (j-6) \right) \right] \quad.
\end{equation}
Numerically, $\upsilon_1 = 0$, $\upsilon_2 = -\upsilon_9 \approx 0.22$, $\upsilon_3 = -\upsilon_8 \approx 0.34$, $\upsilon_4 = -\upsilon_7 \approx 0.30$, and $\upsilon_5 = -\upsilon_6 \approx 0.12$.
We can thus relate experimentally observed shear angle profiles $\gamma(s,t)$ to sliding displacements between adjacent DMTs, with $\Delta_j(s,t)$ denoting the sliding displacement between DMTs $j$ and $j+1$.

For comparison with the continuum description of \cite{SartoriTwist2016},
our discrete sliding displacements satisfy
$\Delta_j \approx - \sin(2\pi/9)\Delta(\phi_{j+1/2})$,
where the azimuthal angle $\phi_{j+1/2}=(2\pi/9)(j-3)$ marks the center between DMTs $j$ and $j+1$.
For twisted, nonplanar axonemal shapes, the simple definition Eq.~\eqref{eq:Delta_j_def} 
should be replaced by a definition that compares arclengths at filament positions 
connected along a direction perpendicular to the local filament tangent~\cite{SartoriTwist2016}; 
this definition results in a twist-induced offset to the sliding displacements, 
see Eq.~\eqref{eq:Delta_j_twist}.

\section{Appendix: Bending- and twist-induced sliding energies decouple}
We show that bending-induced and twist-induced contributions decouple in a simple model of sliding elasticity,
following closely~\cite{SartoriTwist2016}. 
We assume a sliding-elastic energy of the form
\begin{equation}
E_s =
\int_0^L\!ds\,\sum_{j=1}^9 \frac{K_j}{2}\Delta_j^2 \quad,
\label{eq:E_s}
\end{equation}
which models elastic connections between neighboring DMTs as linear springs.
In the presence of twist $\Omega_3$, the relation Eq.~\eqref{eq:Delta_j} acquires an additional contribution $\Delta_\mathrm{twist}$ from twist and reads to linear order~\cite{SartoriTwist2016}
\begin{equation}
\Delta_j = \upsilon_j a\gamma + \Delta_\mathrm{twist}
\quad,
\label{eq:Delta_j_twist}
\end{equation}
where in our discrete description
\begin{equation}
\Delta_\mathrm{twist}
=
\sin\!\left(\frac{2\pi}{9}\right)
\left(\frac{a}{2}\right)^2
\Omega_3
\quad.
\end{equation}
For $K_2=\ldots=K_9$, the cross-term between bending and twist in Eq.~\eqref{eq:E_s} vanishes
because $\upsilon_1=0$ and $\sum_j\upsilon_j=0$.
Thus, the sliding-elastic energy separates into independent contributions from bending and twist,
\begin{equation}
E_s =
\int_0^L\!ds\,
\left[
\frac{K a^2}{2}\gamma^2
+
\frac{C}{2}\Omega_3^2
\right]
\quad,
\end{equation}
where the effective sliding stiffness of the bending mode is
\begin{equation}
K=\sum_{j=1}^9 K_j\upsilon_j^2\quad,
\end{equation}
and the effective twist rigidity due to sliding elasticity is
\begin{equation}
C=
\sin^2\!\left(\frac{2\pi}{9}\right)
\left(\frac{a}{2}\right)^4
\sum_{j=1}^9 K_j
\quad.
\end{equation}
Hence, twist does not modify the bending-associated sliding elasticity to linear order.
Under an analogous assumption of identical linear friction laws for sliding between neighboring DMTs, the bending-associated and twist-induced friction contributions also decouple to linear order.

As in \cite{striegler_twisttorsion_2025}, we define twist by
$\Omega_3 = \e_2\cdot\partial_s\e_1$,
where $(\e_1,\e_2,\e_3)$ denotes a right-handed material frame of the axoneme
with $\e_3=\tvec$, chosen with $\e_1\approx-\n$ and $\e_2\approx-\bvec$.
Our sliding-displacement convention has the opposite sign
to that of Sartori et al.~\cite{SartoriTwist2016}; 
correspondingly the twist-induced contribution to $\Delta_j$ in Eq.~\eqref{eq:Delta_j_twist}
also enters with the opposite sign.

\section{Appendix: Force estimate accounting for twist-induced correction of sliding displacements}
As an additional test on the three-dimensional waveform data from~\cite{striegler_twisttorsion_2025}, 
we define a smaller region-of-trust $\Omega_{\mathrm{trust,3D}}$ 
without assuming identical sliding stiffnesses $K_j$ of the individual DMT pairs. 
We consider only space-time points for which torsion $\tau$ can be reliably determined, and 
assuming twist-torsion coupling $\Omega_3=\tau$, the twist-induced corrections of the sliding displacements $\Delta_j$ can be computed from Eq.~\eqref{eq:Delta_j_twist}.
We further restrict the analysis to space-time points for which the twist-corrected sliding displacements $\Delta_j$ satisfy
\begin{equation}
\gamma\upsilon_j\Delta_j>0 \quad,\quad j=2,\ldots,9 \quad.
\label{eq:stringent_condition_3D}
\end{equation}
This condition ensures that the individual contributions $-a K_j\upsilon_j\Delta_j$
from each DMT pair to the bending-mode sliding-elastic force have the same sign as in the planar case.
Since these contributions all have the same sign in the planar case,
Eq.~\eqref{eq:stringent_condition_3D}
ensures that this remains true when the twist correction is included.
As a consequence, the sum of these contributions has the same sign as assumed in the definition of 
$\Omega_{\mathrm{trust}}$, namely $-\sign(\gamma)$,
irrespective of the individual positive stiffnesses $K_j$.
The condition can be evaluated directly from the measured three-dimensional waveform
without extrapolating torsion into regions of low curvature.
With this stringent restriction, $\Omega_{\mathrm{trust,3D}}$ comprises only $1.2\%$ of all space-time points (11 points).
Within this reduced region, we obtain
$F^\mathrm{95th}_\mathrm{head}=6.5\,\pN$,
while the median of $|F_\mathrm{head}|$ is $3.6\,\pN$.

\section{Appendix: Motor distribution minimizing the inferred force per head}
As an extreme case for the distribution of dynein motors on the doublet microtubules, 
we can assign all dynein motors to the doublet microtubules with the absolutely largest geometric prefactors, $j=3$ and $j=8$, 
and assume that all motors on one side of the axoneme are active at a given time, while all motors on the opposite side are inactive. 
Accordingly, for $f_m(s,t)>0$, we assign $n_3=1$ and $n_j=0$ for $j\neq 3$, 
whereas for $f_m(s,t)<0$, we assign $n_8=1$ and $n_j=0$ for $j\neq 8$. 
If $\rho=N/L$ denotes the total line density of dynein motor heads in the axoneme, this assumption assigns a line density $\rho/2$ to the active side. 
The inferred force per active motor head therefore obeys, using $\upsilon_\text{max} = |\upsilon_3|=|\upsilon_8|$,
\begin{equation}
\frac{|F_\mathrm{passive,known}|}{a\upsilon_\mathrm{max}\,\rho/2}
< \max_j F_j 
\label{eq:F_head_conservative}
\end{equation}
Again, this lower bound applies to all space-time points within the region-of-trust.
We thus obtain lower bounds on the peak force per active dynein head analogous to those reported in Fig.~\ref{fig3}; in this limiting case, however, the bounds are reduced by the factor $[\upsilon_\mathrm{max}(\rho/2)/\rhoDMT]^{-1}\approx 0.74$ relative to the more realistic motor distribution assumed in Fig.~\ref{fig3}.

\begin{table*}
\centering
\scriptsize
\resizebox{0.8\textwidth}{!}{%
\begin{tabular}{|l|c|rrrrrrr|rrr|rrr|r|r|r||r|}
\hline
Region & DMT &
\multicolumn{7}{|c}{IDA} & 
\multicolumn{3}{|c}{MD} & 
\multicolumn{3}{|c|}{ODA} & 
IDA/ & ODA & Total & Total/$\micron$ \\
& & 
f & a & b & c & e & g & d & 
1 & 2 & 3 & 
$\alpha$ & $\beta$ & $\gamma$  
& \multicolumn{1}{|c|}{MD} & & & \\
\hline\hline
proximal & 1
& 0 & 0 & 0 & 0 & 0 & 1 & 0
& 1 & 1 & 1
& 4 & 4 & 4
& 4 & 12 & 16 & 166.7 \\
\hline
proximal & 2
& 1 & 1 & 0 & 1 & 1 & 1 & 0
& 0 & 0 & 1
& 4 & 4 & 4
& 7 & 12 & 19 & 197.9 \\
\hline
proximal & 3
& 1 & 1 & 0 & 1 & 1 & 1 & 0
& 0 & 0 & 1
& 4 & 4 & 4
& 7 & 12 & 19 & 197.9 \\
\hline
proximal & 4
& 1 & 1 & 0 & 1 & 1 & 1 & 0
& 0 & 0 & 1
& 4 & 4 & 4
& 7 & 12 & 19 & 197.9 \\
\hline
proximal & 5
& 1 & 1 & 0 & 1 & 1 & 1 & 0
& 0 & 0 & 1
& 4 & 4 & 4
& 7 & 12 & 19 & 197.9 \\
\hline
proximal & 6
& 1 & 1 & 0 & 1 & 1 & 1 & 0
& 0 & 0 & 1
& 4 & 4 & 4
& 7 & 12 & 19 & 197.9 \\
\hline
proximal & 7
& 1 & 1 & 0 & 1 & 1 & 1 & 0
& 0 & 0 & 1
& 4 & 4 & 4
& 7 & 12 & 19 & 197.9 \\
\hline
proximal & 8
& 1 & 1 & 0 & 1 & 1 & 1 & 0
& 0 & 0 & 1
& 4 & 4 & 4
& 7 & 12 & 19 & 197.9 \\
\hline
proximal & 9
& 1 & 1 & 0 & 1 & 1 & 1 & 0
& 0 & 0 & 1
& 4 & 4 & 4
& 7 & 12 & 19 & 197.9 \\
\hline\hline
centr./dist. & 1
& 1 & 1 & 0 & 0 & 0 & 1 & 1
& 1 & 0 & 0
& 4 & 4 & 4
& 6 & 12 & 18 & 187.5 \\
\hline
centr./dist. & 2
& 1 & 1 & 1 & 1 & 1 & 1 & 1
& 0 & 0 & 0
& 4 & 4 & 4
& 8 & 12 & 20 & 208.3 \\
\hline
centr./dist. & 3
& 1 & 1 & 1 & 1 & 1 & 1 & 1
& 0 & 0 & 0
& 4 & 4 & 4
& 8 & 12 & 20 & 208.3 \\
\hline
centr./dist. & 4
& 1 & 1 & 1 & 1 & 1 & 1 & 1
& 0 & 0 & 0
& 4 & 4 & 4
& 8 & 12 & 20 & 208.3 \\
\hline
centr./dist. & 5
& 1 & 1 & 1 & 1 & 1 & 1 & 1
& 0 & 0 & 0
& 4 & 4 & 4
& 8 & 12 & 20 & 208.3 \\
\hline
centr./dist. & 6
& 1 & 1 & 1 & 1 & 1 & 1 & 1
& 0 & 0 & 0
& 4 & 4 & 4
& 8 & 12 & 20 & 208.3 \\
\hline
centr./dist. & 7
& 1 & 1 & 1 & 1 & 1 & 1 & 1
& 0 & 0 & 0
& 4 & 4 & 4
& 8 & 12 & 20 & 208.3 \\
\hline
centr./dist. & 8
& 1 & 1 & 1 & 1 & 1 & 1 & 1
& 0 & 0 & 0
& 4 & 4 & 4
& 8 & 12 & 20 & 208.3 \\
\hline
centr./dist. & 9
& 1 & 1 & 0 & 1 & 1 & 1 & 1
& 0 & 0 & 0
& 4 & 4 & 4
& 7 & 12 & 19 & 197.9 \\
\hline
\end{tabular}%
}
\caption[]{
Approximate dynein-head counts per DMT and 96-$\nm$ repeat inferred from Bui~et~al.~\cite{bui_polarity_2012}. 
Dynein-f is counted as two heads; other IADs and minor dyneins are counted as one head. 
Each ODA heavy chain is counted with four copies per 96-$\nm$ repeat.
}
\label{table:Bui}
\end{table*}

\end{document}